\documentclass[sn-aps]{sn-jnl}% referee option is meant for double line spacing

\usepackage{multirow}%
\usepackage{amsmath,amssymb,amsfonts}%
\usepackage{amsthm}%
\usepackage{mathrsfs}%
\usepackage[title]{appendix}%
\usepackage{xcolor}%
\usepackage{textcomp}%
\usepackage{manyfoot}%
\usepackage{booktabs}%
\usepackage{listings}%

\newcommand{\p}{\partial}
\newcommand{\mX}{\mathbb{X}}
\newcommand{\mY}{\mathbb{Y}}

\newcommand{\vep}{\varepsilon}
\newcommand{\R}{\mathbb{R}}

\newcommand{\pol}{V(x)}
\newcommand{\dd}{\mathrm{d}}
\newcommand{\expp}[2]{\exp\left( \frac{#1}{#2}\right)}
\newcommand{\inner}[2]{\left\langle #1,#2 \right\rangle}
\newcommand{\transp}{\mathrm{T}}

\begin{document}
	
\title{Quasi-potential and drift decomposition in stochastic systems by sparse identification}

\author[1]{\fnm{Leonardo} \sur{Grigorio}}\email{leonardo.grigorio@cefet-rj.br}

\author[2]{\fnm{Mnerh} \sur{Alqahtani}}\email{mnalqahtani@uhb.edu.sa}

\affil[1]{\orgname{Centro Federal de Educação Tecnológica Celso Suckow da Fonseca}, \orgaddress{\city{Nova Friburgo}, \postcode{28635-000}, \state{RJ}, \country{Brazil}}}

\affil[2]{\orgdiv{Mathematics department}, \orgname{University of Hafr Albatin}, \orgaddress{\city{Hafar Al Batin}, \postcode{39524}, \country{Saudi Arabia}}}

\abstract{
The quasi-potential is a key concept in stochastic systems as it accounts for the long-term behavior of the dynamics of such systems. It also allows us to estimate mean exit times from the attractors of the system, and transition rates between states. This is of significance in many applications across various areas such as physics, biology, ecology, and economy. Computation of the quasi-potential is often obtained via a functional minimization problem that can be challenging. This paper combines a sparse learning technique with action minimization methods in order to: (i) Identify the orthogonal decomposition of the deterministic vector field (drift) driving the stochastic dynamics; (ii) Determine the quasi-potential from this decomposition. This decomposition of the drift vector field into its gradient and orthogonal parts is accomplished with the help of a machine learning-based sparse identification technique. Specifically, the so-called \emph{sparse identification of non-linear dynamics} (SINDy) \cite{BPK:2016} is applied to the most likely trajectory in a stochastic system (\emph{instanton}) to learn the orthogonal decomposition of the drift. Consequently, the quasi-potential can be evaluated even at points outside the instanton path, allowing our method to provide the complete quasi-potential landscape from this single trajectory. Additionally, the orthogonal drift component obtained within our framework is important as a correction to the exponential decay of transition rates and exit times. We implemented the proposed approach in 2- and 3-D systems, covering various types of potential landscapes and attractors.}


\maketitle
\section{Introduction}
% The introduction contains 6 (6 paragraphs):

% 1) Background: related between LDT & quasi-potential, and the importance of quasi-potential is the literature. 

In the large deviation theory, the leading order of the probability of long-term behavior of the system under consideration is given asymptotically by the \emph{rate function} (Aka. the \emph{action}) minimum. On the other hand, the minimum of the rate function over trajectory space, achieving final-time configuration on the first time hitting $T$, is equivalent to the quasi-potential $\phi:\R^d \rightarrow \R^d$ \cite{dembo-zeitouni:2010, grafke-vanden-eijnden:2019}, that is,
\begin{equation}
\label{eq:Phi_S}
\phi(x) = \inf_{T>0} \, \min_{\varphi \in A_x} S[\varphi],
\end{equation}
where, 
\begin{equation}
\label{eq:A_x}
A_x = \{ \varphi \in \text{AC}_{[0, \, T]} \left(\R^d \right)  | \varphi(0) = \bar{x}, \, \varphi(T) = x \},
\end{equation}
and $\bar{x}$ is a stable attractor. The space $\text{AC}_{[0, \, T]} \left(\R^d \right)$ is the space of absolutely continuous functions $\varphi : [0, \, T] \rightarrow \R^d$.
%Here, the quasi-potential is defined as the minimum action required for the system to transition from one state to another.
This highlights the importance of the quasi-potential $\phi(x)$ in predicting the asymptotic behavior of a system subject to random fluctuations.
%The quasi-potential has a pivotal role in describing stochastic systems as it accounts for the long-time behavior of the dynamics of such systems.
For example, it is relevant in studying climate and atmosphere dynamics \cite{galfi:2021, ragone_etal:2018,bouchet_nardini_tangarife:2013}, as well as in quantifying stability in ecological models \cite{NA:2016} and the dynamics of multi-cellular organisms \cite{Zhou_etal:2012}. Furthermore, it is closely linked to rare events \cite{bouchet_etal:2019}, providing estimates for the mean exit times from the attractor and also transition probabilities \cite{TOU:2009}.

% 2) Review of the traditional methods of computing the quasi-potential & the associated challenges. 

Evaluating quasi-potential, especially for non-linear and high-dimensional systems, is a difficult task. Overall, there are two main approaches to calculating quasi-potentials \cite{Lin_etal:2021}. One is rooted in the variational formulation of the Freidlin-Wentzell action \cite{freidlin-wentzell:2012}. Aside from the computation of the quasi-potential, this method provides the path connecting the states, known as the \emph{instanton} or \emph{minimum action path}, and is thus referred to as the instanton method \cite{grafke-grauer-schaefer:2015}. However, this method does not produce the quasi-potential outside the transition paths. The second approach involves solving the static Hamilton-Jacobi (HJ), a non-linear partial differential equation, which is typically arduous to solve as standard methods like finite differences or finite elements often fail. Unlike the path-based approach, this method not only yields a local quasi-potential along the path but also generates a global landscape by solving the HJ equation around different attractors, and then combining them into a global quasi-potential by imposing continuity at the saddle points. This method requires dedicated algorithms to solve the HJ equation on a mesh using an enhanced version of ordered upwind methods \cite{CAMERON:2012,Cameron_etal:2019}.

% 3) the new methods of computing the quasi-potential based on ML & the associated challenges. 

In recent years, research has been introducing methods to tackle quasi-potential calculation incorporating machine learning techniques. This involves the use of neural networks for stochastic differential equations with different types of noise, such as additive \cite{Lin_etal:2021}, multiplicative, and non-Gaussian noise \cite{Li_etal:2021,Li_etal:2022}. Following a more analytically oriented route, \cite{Bouchet_etal:2016} develops a perturbative calculation, providing a power series expression for the quasi-potential. Common to many of the preceding approaches is the use of an orthogonal decomposition of the deterministic vector field (the drift) of the system; one component is the gradient of a scalar function and the other is normal to the first component. This is central to many methods of evaluating the quasi-potential as the quasi-potential can be directly computed if this decomposition is available \cite{Zhou_etal:2012,Cameron_Yang:2019}.

%[{\color{red} Shall we keep this paragraph? Maybe replace it in the discussion?} In a sense, the quasi-potential is an extension of the concept of potential in physics. When a particle is subject to a potential energy, a force develops over the particle pointing to the descent direction of the potential landscape. In the presence of dissipation, the minimum of the potential becomes an attractor of the dynamics. To leave this attractor, an amount of work is required to climb the potential. Even when the force cannot be derived from a potential, it is still possible, in some circumstances, to decompose it as a sum of a gradient plus an orthogonal component. The orthogonal force contributes to the movement of the particle parallel to the iso-lines (contour-line) of the potential. There is no work required to move the system along these level curves. ]

%In the context of weak noise-driven dynamical systems, the drift (deterministic) term plays the role of a \textit{“force”}. When the drift is a gradient function, the potential alone can determine the exponential scaling of the probability distribution of finding the system drifts away from the attractor. In situations when the system poses a non-gradient drift, \texttt{\color{blue} [I need to continue from here]} it has a component that is written as a gradient of a function $\phi$, the probability distribution decays exponentially with scaling being determined by $\phi$, called quasi-potential or rate function \cite{freidlin-wentzell:2012,TOU:2009}. 

% 4) Independent paragraph about SINDy

In the midst of machine learning-like tools, non-linear system identification is a method for discovering the underlying structure of a non-linear dynamical system from data \cite{bongard:2007, schmidt:2009}. It employs symbolic regression \cite{koza:1992} to compute the non-linear functions of the variables and their temporal derivatives. However, relying just on symbolic regression can be computationally costly and susceptible to overfitting; hence, sparse representation makes use of the fact that most physical systems contain only a few non-linear variables, creating \emph{sparse} matrices \cite{BPK:2016}.
 
% 5) The main results of the paper
    
In this study, we will combine a sparse learning technique with the instanton method to uncover the underlying structure of the drift. More concretely, we will adapt sparse identification of non-linear dynamics (SINDy) to use the instanton path as input in order to identify the orthogonal decomposition of the drift, which in turn allows for the determination of the quasi-potential in analytical form. The advantage of this approach is that a global quasi-potential can be computed from a single trajectory, even at points lying outside the trajectory. Typically, obtaining the quasi-potential in a given domain within the instanton framework requires numerous realizations, which can be prohibitive in systems with high dimensionality and non-linearity. Besides, our method also yields the normal component of the drift, which is important for the evaluation of sub-exponential prefactors that correct the decay rate of statistical quantities as transition probabilities, first exit times, and mean first passage times \cite{BR:2022, vezzani:2024, grafke:2024}.

% 6) The paper instruction: 

 The paper is organized as follows: Sections \ref{sec:Inst} and \ref{sec:Sparse} provide theoretical and numerical background on the instanton's role in drift decomposition and sparse identification of the system under consideration, respectively. In particular, the general form of SINDy (see section \ref{sec:SINDy}) will be explained in terms of instanton, with the output of instanton solutions used as an input of SINDy in section \ref{sec:SINDy_Inst} to find the gradient of the quasi-potential and the normal to the gradient component (components of the drift decomposition). In Section \ref{sec:results}, the proposed approach will be applied to and tested on several systems of multiple dimensions. Finally, Section \ref{Sec:Conclusion} discusses the limits and prospective extensions.
    
\section{Instanton and drift decomposition}
\label{sec:Inst}
Consider the diffusion process defined by
\begin{equation}
\label{eqsde1}
\dd X_t = b(X_t)\dd t + \sqrt{\vep} \sigma \dd W_t\,, \quad X_0 = \bar{x}, 
\end{equation}
with a deterministic drift $b:\R^d \rightarrow \R^d$. The term $\dd W_t$ denotes a Wiener increment, $\vep$ is the noise amplitude, and the diffusion matrix prescribed by the noise covariance $a = \sigma \sigma^{\transp}$ with $\sigma \in \R^{d}$. 

In such stochastic systems, the asymptotic estimate of the probability of rare events of interest, e.g., state transition rate and realizing extreme events, is provided by the large deviation principle \cite{dembo-zeitouni:2010, ellis-etal:1984, freidlin-wentzell:2012}. It states that the probability of a set of rare events in the limit of vanishing $\vep$ is dominated by the minimizer of the rate function (action), 
%\begin{equation}
%\label{eq1}
%p(x,t|x_0) = A \expp{-\inf_{\varphi} S[\varphi]}{\vep}, %   \frac{-\inf_{\varphi} S[\varphi]}{\vep}} ,
%\end{equation}
\begin{equation}
\label{eq1}  
p(x, T \, | \, \bar{x}) = A \expp{-\inf_{\varphi} S[\varphi (t)]}{\vep}, %   \frac{-\inf_{\varphi} S[\varphi]}{\vep}} ,
\end{equation}
where $p(x, T \, | \, \bar{x})$ represents the probability measure of $\varphi$ reaching $x$ for the first time at $T$, i.e.  $\varphi \in A_x$, as in equation (\ref{eq:A_x}), and $A$ is a normalization constant. The action functional $S[\varphi]$ is given through \cite{freidlin-wentzell:2012},
\begin{equation}
\label{eq1a}
S[\varphi(t)] = \frac{1}{2}\int_0^{T} \inner{\dot \varphi - b}{a^{-1}(\dot \varphi - b)}  \dd t.
\end{equation}
Now, the optimization problem $\inf_{\varphi} S[\varphi (t)]$ needs to be solved to estimate the probability (\ref{eq1}).

For this purpose, the instanton equations can be formulated using the Hamiltonian formalism; thus, it is called as well Hamiliton’s equations. The solution of the instanton equations is the minimizer of the action $S[\varphi (t)]$. First, the Hamiltonian defined as
the Legendre-Fenchel (L-F) transform of the Lagrangian, that is,
\begin{equation}
\label{eq:Hamilton}
H(\varphi,\vartheta) = \sup_{\dot{\varphi}}\left( \inner{\vartheta}{\dot \varphi}-L(\varphi,\dot \varphi)\right),
\end{equation}
where $\vartheta = \p L/\p \dot{\varphi}$ is the conjugate momenta of $\varphi$, and the Lagrangian is,
\begin{equation}
\label{eq:Lagrangian}
 L(\varphi,\dot \varphi) = \frac 12 \inner{\dot\varphi - b(\varphi)}{a^{-1}(\dot\varphi - b(\varphi))},
\end{equation}
which is merely the integrand of $S[\varphi (t)]$, see equation (\ref{eq1a}). Notice that the Lagrangian can be written as the L-F  transform of the Hamiltonian as well, i.e., ${L(\varphi,\dot \varphi)  = \sup_{\vartheta}\left( \inner{\vartheta}{\dot \varphi}- H(\varphi,\vartheta)\right).}$ The Hamiltonian (\ref{eq:Hamilton}) after performing the optimization is just
\begin{equation}
\label{eq:H}
H(\varphi,\vartheta) = \frac 12 \inner{\vartheta}{a  \, \vartheta}+\inner{b}{\vartheta}.
\end{equation}
When minimizing (\ref{eq1a}) using Hamiltonian-Lagrangian information, it leads to a set of instanton equations,
\begin{align}
\label{eq:hamiltons_a}
\dot{\varphi} =\, & \p_\vartheta H =  b(\varphi) +  a\vartheta \, ,\\
\label{eq:hamiltons_b}
\dot{\vartheta} = & -\p_\varphi H = -  (\p_\varphi b)^{\mathrm{T}} \vartheta\,.
\end{align}
Solving equations (\ref{eq:hamiltons_a}-\ref{eq:hamiltons_b}) with appropriate initial/final conditions allows us to find the most likely trajectory. This method constitutes the ground for many instanton and quasi-potential calculations \cite{GSE:2017, schorlepp:2023}, and their applications \cite{schorlepp:2022, apolinario:2022, alqahtani:2022}. 

% Hereafter we consider 
% \begin{equation}
% \label{eq2}
% \varphi(T) = x, \;\; \varphi(0) = x_0 \in B
% \end{equation} 

Now, we seek to rewrite the temporal evolution of the instanton $\varphi$ (\ref{eq:hamiltons_a}) in terms of the drift decomposition. Let $V(x)$ denote the potential of the system (\ref{eqsde1}). Combining equations (\ref{eq1}), (\ref{eq:Phi_S}), and $V(x) = \phi(x)/2$ \cite{freidlin-wentzell:2012} results in the probability measure in terms of the potential,
\begin{equation}
\label{eq3}
p(x,T|\bar{x}) = A \expp{-2 \, \pol}{\vep},
\end{equation}
which in turn can be expressed as
\begin{align}
\label{eq4}
p(x,T|\bar{x}) & =  A \exp\left(-\frac{2}{\vep}\int_0^{T} \frac{\dd V}{\dd t}\,\dd t\right) \nonumber \\
 & = A\exp\left(-\frac{2}{\vep}\int_0^{T}  \left( \, \p_t V + \inner{\nabla V}{\dot \varphi}  \right) \dd t \right).
\end{align}
By comparing equations (\ref{eq1}) and (\ref{eq1a}) with (\ref{eq4}), we find that,
\begin{equation}
\label{eq5}
\begin{split} 
- & \frac{1}{2\vep} \int_0^{T} \inner{\dot \varphi - b}{a^{-1}(\dot \varphi - b)} \,\dd t \\
= &-\frac{2}{\vep} \int_0^{T} \left( \, \p_t V + \inner{\nabla V}{\dot \varphi}  \right) \, \dd t \,.
\end{split} 
\end{equation}
Omitting the integral symbol, the above expression becomes,
\begin{align}
\label{eq6}
\frac{1}{4} \inner{\dot \varphi}{a^{-1}\dot \varphi} + \frac{1}{4} \inner{ b}{a^{-1} b} - \frac{1}{2} \inner{\dot \varphi}{a^{-1} b} - \inner{\dot \varphi}{\nabla V} = \p_t V.
\end{align}
Using the decomposition (refer to equation (\ref{A4a}) in Appendix \ref{sec:app})
\begin{equation}
\label{eq:drift}
b(x) = - a \nabla V + l(x),
\end{equation}
into (\ref{eq6}), yields 
\begin{align}
\label{eq8}
\frac{1}{4} \inner{\dot \varphi}{a^{-1}\dot \varphi} + \frac{1}{4} \inner{a \nabla V}{\nabla V} + \frac{1}{4}\inner{l}{a^{-1}l} + \nonumber \\
- \frac{1}{2}\inner{\nabla V}{l}- \inner{\dot \varphi}{\nabla V +a^{-1}l} = \p_t V\,,%\underbrace{\p_t V\,,}%_{\hspace{14pt} -\inner{\nabla V}{l} \; \mathrm{,see (\ref{A6})}}
\end{align}
where $\p_t V = -\inner{\nabla V}{l}$, see equation (\ref{A6}) for details. Throughout this paper, we are assuming stationarity, $\p_t V = 0$, implying the orthogonality condition, 
\begin{equation}
\label{eq:orthogonal}
\inner{\nabla V}{l} = 0\,.
\end{equation}
Finally, equation (\ref{eq8}) is rewritten with the integral symbol restored as,
\begin{equation}
\label{eq9}
\frac{1}{2} \int_0^{T} \inner{\dot \varphi- a \nabla V -l}{a^{-1}\left(\dot \varphi- a \nabla V -l\right)} \dd t = 0 \,.
\end{equation}
The integrand is non-negative, so the instanton satisfies,
\begin{equation}
\label{eq:instanton}
\dot \varphi = a \nabla V + l.
\end{equation}
Note that if $a = I$, i.e., the identity, equation (\ref{eq:instanton}) aligns with previous results \cite{applied:2019,freidlin-wentzell:2012,CAMERON:2012,BR:2022}. In the sequel, we consider $a = I$. Equation (\ref{eq:instanton}) can be interpreted as follows: The instanton follows the rotational component of the field but goes against its potential component \cite{CAMERON:2012}. In the absence of the rotational component $l$, (\ref{eq:instanton}) indicates that instanton obeys,
\begin{equation}
\label{eq:gradient}
\dot \varphi = \nabla V(\varphi).
\end{equation}
This shows that in order for the system to leave the attractor $\varphi(0) = \bar{x}$, it must overcome the potential barrier $V(x) = \phi(x)/2$. It is in this sense that the quasi-potential $\phi(x)$ generalizes the concept of potential to non-gradient systems since $l(x)$ is transverse to $\nabla \phi(x)$. Moreover, equation (\ref{eq:gradient}) is also understood as a time-reversed dynamics, as opposed to the deterministic dynamics $\dot \varphi = -\nabla V $.  Hence, the rotational component $l$ disrupts the time-reversal symmetry \cite{bouchet_tangarife_etal:2014}.

Now, we are in a position to establish the equations that
are at the core of our paper. Combining the drift decomposition (\ref{eq:drift}) together with both instanton equations (\ref{eq:hamiltons_a}) and (\ref{eq:instanton}) results in 
\begin{equation}
\label{eq:p_grad}
\vartheta(t) = \nabla \phi(\varphi(t)) = 2 \nabla V(\varphi(t)) \,.
\end{equation}
There are a few points that need to be highlighted, 
\begin{itemize}
\item The conjugate momentum $\vartheta$ corresponds to the gradient component of the drift. This finding is consistent with results from large deviation theory, where $\vartheta(T)$ is the gradient of the rate function in the G\"artner-Ellis theorem \cite{ellis-etal:1984}. It is also in line with the general Hamilton-Jacobi theory of mechanics \cite{lanczos:1949}, where the conjugate momentum is the gradient of the action. This equation has been presented also in ref. \cite{Bouchet_etal:2016}. Here, we remark that the relationship holds for all time $t\in[0,T]$, i.e., along the trajectory, not just at the final time, which we will exploit later.

\item $\vartheta$ encompasses all the information about the quasi-potential $\phi$ in the sense that $\phi$ can be obtained from $\vartheta$ as will be discussed along the text. 

\item Note that in equation (\ref{eq:hamiltons_a}), one can understand the dynamics of the instanton as being subject to two forces. The first term on the right-hand side is the deterministic drift, driving the system regardless of the noise. The second is regarded as the optimal force required to take the system to a given state. Specifically, from (\ref{eq:hamiltons_a}) and (\ref{eq:p_grad}), we have $\dot{\varphi} = b + 2 a\nabla V$, indicating that the contribution of the second term is of a gradient type, hence helping the system to surpass potential barriers.

\item The case $\nabla \phi = 0$ prevents the decomposition given by equation (\ref{eq:drift}). This means that $\vartheta = 0$, indicating no force pushing the system against the potential. Further, instanton satisfies $\dot{\varphi} = b$, according to (\ref{eq:hamiltons_a}), implying the system follows the deterministic dynamics with vanishing action. In this situation, there is no work to be realized by the noise. The system relaxes towards the attractor. 
\end{itemize}

For the orthogonal component of the drift, we combine equations (\ref{eq:instanton}) and (\ref{eq:p_grad}) to find
\begin{equation}
\label{eq:orthogonal_comp}
l(t) = \dot{\varphi}(t)- \frac 12 a\vartheta(t) \, .
\end{equation}
Equations (\ref{eq:p_grad}-\ref{eq:orthogonal_comp}) relate the decomposition of the drift in terms of the minimum action path (instanton) $\varphi(t)$ and its conjugate momentum $\vartheta(t)$.

However, as far as solutions of Hamilton’s equations are concerned, equations (\ref{eq:p_grad}-\ref{eq:orthogonal_comp}) do not provide the functional dependence of the decomposition ${b(x) = - \nabla V(x) +l(x)}$ outside the instanton trajectory $\varphi(t)$. Therefore, the problem we aim to tackle is: How to obtain the drift components, $\nabla V(x)$ and $l(x)$, across the domain $x \in D \subset \R^d$ from a single pair of trajectories  $(\vartheta(t),\varphi(t) )$? Next, we will demonstrate that SINDy can effectively answer this question.

\section{Sparse identification}
\label{sec:Sparse}

The construction of mathematical models from data presents a fundamental challenge across numerous scientific and engineering domains. Sparse Identification of Nonlinear Dynamics (SINDy)  \cite{BPK:2016} is an interpretable machine learning-based approach utilized for the discovery of governing equations of dynamical systems. The primary objective of SINDy is to deduce the underlying governing equation from data by finding key terms that characterize the dynamics of complex systems. In this way, it is a data-driven approach to identifying potential correlations among dynamical variables.

Section \ref{sec:SINDy} presents a general and summarised review of the SINDy technique. In Section \ref{sec:SINDy_Inst}, we explore its use from the perspective of instanton knowledge.

\subsection{SINDy overview}
\label{sec:SINDy}

Designed to learn governing equations of dynamical systems from data measurements, SINDy has been applied to non-linear oscillators, reduced-dimensional fluid models \citep{BPK:2016,Loiseau_Brunton_2018}, the determination of Kramers-Moyal coefficients \cite{BFC:2018,callaham:2021}, non-linear optics \cite{Sorokina:16}, plasma physics \cite{Dam_etal:2017}, and partial differential equations \cite{Rudy_etal:2017}, among other areas. Following the notation of \cite{BFC:2018}, we briefly review the rationale behind SINDy below.

Let $x, \, y \in \R^d$ and given  time series of data,  $y_i(t_1),y_i(t_2),\dots,y_i(t_N)$ and $x_i(t_1),x_i(t_2),\dots,x_i(t_N)$ with $i=1,\dots,d$. Further, assume that there exists a relation $y = F(x)$ between $x$ and $y$. According to SINDy, the problem of finding $F$ is handled as follows. First, we organize $y$ data into a matrix $\mY_{N \times d}$,
\begin{equation}
\label{eq:Y}
 \mY = \left[
\begin{array}{cccc}
| &| &  & |  \\
y_1 & y_2 & \dots & y_d \\
| & |& &|
\end{array}\right],
\end{equation}
such that each row contains a snapshot, that is, ${Y_{ij} = y_j(t_i)}$.
SINDy basically amounts to represent $\mY$ as a linear combination of pre-selected basis functions 
\begin{equation}
\label{eq:YXc}
\mathbb{Y} = \mathbb{X} \, c \, ,  
\end{equation}
where $\mX_{N \times K}$ is a library matrix in which each column represents a user-defined basis functions
\begin{equation}
\label{eq:library}
\mX = 
\left[
  \begin{array}{ccccccccc}
     | & | & | & | & |& |& | &|&|    \\
     1 & x_1 & x_2 & x_1^2 & x_2^2 & x_1x_2 & ...  & \cos(x_1) & \dots \\
    | & | & | &|&|&|&| & | &|
  \end{array}
\right],
\end{equation}
and $c_{K\times d}$ is an unknown coefficients matrix,
 \begin{equation}
\label{eq:Y_}
 c = \left[
\begin{array}{cccc}
| &| &  & |  \\
\mathbf{c}_1 & \mathbf{c}_2 & \dots & \mathbf{c}_d \\
| & |& &|
\end{array}\right]\,,
\end{equation}
with $\mathbf{c}_j \in \R^K$, $j=1,\dots,d$.
Each element of $\mathbf{c}_j$ gives the relative weight of the candidate function to the $j^{\text{th}}$ component of $y$.
 
There is a great deal of freedom in constructing $\mX$. As a rule, the choice should be guided by the nature and symmetries of the problem at hand. 
%The problem (\ref{eq:YXc}) is understood to be solved in the least-square sense since, in most practical cases, $N >> K$. 
In practical cases where $N >> K$, problem (\ref{eq:YXc}) is typically solved using the least-square sense. We seek, thus, a minimizer $\tilde{c}$ that satisfies,
\begin{equation}
\label{eq:ls}
\tilde{c_i} = \underset{c_i \in \R^K}{\mathrm{argmin}}\; (y_i-\mX \mathbf{c}_i)^{\transp}(y_i-\mX \mathbf{c}_i)\,, \;i=1,\dots,d.
\end{equation} 

It is generally observed that $\tilde{c}$ will not be sparse. However, \cite{BPK:2016} observes that for many systems of interest, only a few terms among all conceivable functions are relevant. Therefore, a sparse solution for (\ref{eq:ls}) is intended. One method to enforce sparsity is to modify (\ref{eq:ls}) by penalizing the $L_1$ norm \cite{kutz2013data}. An alternative approach proposed by Ref. \cite{BPK:2016} is to iteratively perform standard linear regression, removing at each step coefficients $c_{ij}, i=1,\dots, K, \, j=1,\dots,d,$ that are smaller than a threshold value $\lambda$. This method is referred to as sequential threshold least-squares (STLS). However, a disadvantage of this method is the introduction of an arbitrary parameter $\lambda$ that controls the level of sparsity.

Alternatively, reference \cite{BFC:2018} adjusts the original SINDy method of enforcing sparsity by sequentially setting the smallest coefficient $\mathbf{c}_{j}$ to zero at each linear regression step. This modification eliminates the need for the threshold parameter $\lambda$. Furthermore, this method uses cross-validation \cite{statisticallearning} to select the best level of sparsity in the least-squares sense, known as the Stepwise Sparse Regressor (SSR). We implemented this approach when applying SINDy in the next section; see Appendix  \ref{sec:app_2} for details.

Originally, SINDy was designed to identify the governing equations of a non-linear dynamical system $dx(t)/dt = F(x)$. In this framework, $\mY$ is constructed from time derivatives of $x_i(t), \; i=1,\dots,d$, such that $F(x)$ is learned from the method.  In our study, we will adapt the SINDy to uncover the orthogonal drift decomposition aided by instanton equations, which will be discussed in the following section.

\subsection{SINDy applied to instanton dynamics}
\label{sec:SINDy_Inst}

As discussed in section \ref{sec:Inst}, the instanton approach renders the stochastic problem of finding the most likely pathway (instanton) for a given fluctuation into a deterministic one. Instanton is determined by an optimization problem, where it minimizes the action functional (\ref{eq1}). This paper aims to use this path to find the drift’s decomposition in terms of its gradient and orthogonal components (\ref{eq:drift}). To achieve this, we apply the SINDy algorithm, using solutions of instanton equations as input.

Specifically, we will work with equations (\ref{eq:p_grad}-\ref{eq:orthogonal_comp}), accounting for the gradient and orthogonal components, reproduced here for convenience
\begin{align}
\label{eq:sindy_A1}
\vartheta(t) = \nabla(\phi(\varphi(t))) = 2 \nabla V(\varphi(t)) \,, \\
\label{eq:sindy_A2}
\dot \varphi(t) - \frac 12 a\vartheta(t) =l(\varphi(t)) \,.
\end{align}
The aim is to determine the right-hand side of equations (\ref{eq:sindy_A1}-\ref{eq:sindy_A2}) as functions of $x \notin \varphi$, which means at points outside of the instanton path. With this, we can now establish the procedure.

First, we will implement the methodology for determining the gradient of the quasi-potential (\ref{eq:sindy_A1}). The procedure steps are as follows:
\begin{enumerate}
\item Numerically solve (\ref{eq:hamiltons_a}- \ref{eq:hamiltons_b}) to find $\varphi(t_n)$ and $\vartheta(t_n)$ for $n=1,\dots, N$, using a suitable method.
\item Set $[\vartheta_i(t_1) \,\dots \,\vartheta_i(t_N)]^{\transp}$ as the $i^{\text{th}}$ column vector to build $\mY$, see (\ref{eq:Y}), where $i$ spans $1,\dots,d$.
\item Choose $f_1,\dots,f_K$ functions to build the library matrix $\mX$, placing $[f_j(\varphi(t_1)) \, \dots \, f_j(\varphi(t_N))]^{\transp}$ as the $j^{\text{th}}$ column of $\mX$ (see \ref{eq:library}).
\item Determine the sparse solution $\tilde{c}$ of coefficients (\ref{eq:ls}).
\end{enumerate}
The sparse solution $\tilde{c}$ identifies which functions are active, selecting the functional form that accurately represents the vector field component.

In a similar way, for the second equation (\ref{eq:sindy_A2}), we seek to construct on the orthogonal component of the drift. The process remains mostly the same with a small modification in step 2: Replace $[\vartheta_i(t_1) \,\dots \,\vartheta_i(t_N)]^{\transp}$ with $[\dot \varphi(t_1) - \frac 12 a\vartheta(t_1)\, \dots \, \dot \varphi(t_N) - \frac 12 a\vartheta(t_N)]^{\transp}$ and proceed with the remaining steps.

Noteworthy, here are a few comments regarding the procedure steps, 
 \begin{itemize} 	
	\item For step 1, the specific numerical method used may vary. For example, if the system has a single fixed point or a single limit cycle, algorithms of Chernykh-Stepanov type \cite{chernykh-stepanov:2001,grafke-grauer-schaefer:2015} can be implemented to solve Hamilton's equations. However, if there is more than one attractor present, the convexity of the quasi-potential changes, requiring different techniques to obtain $\varphi$ and $\vartheta$. One such technique is the geometric minimum action method (GMAM)  \cite{gmam,gmam2,GSE:2017}, which has evolved from the string method and its variants \cite{string:2002,ERV:2007}. 
		
	\item Concerning step 3, the choice of the candidate functions within the SINDy framework is arbitrary but crucial to accurately representing the process. Here, in the context of finding the components of the drift, the selection of functions is naturally guided by the form of the drift. For instance, if the drift consists of polynomial terms, it makes sense to consider the candidate functions $f_1,\dots,f_K$ as polynomials.
	
	\item Regarding step 4, we utilized the SSR \citep{BFC:2018} variation of SINDy, discussed in Appendix \ref{sec:app_2}.
  \end{itemize}

\section{Applications and results}
\label{sec:results}

 This section will exemplify the outlined approach across a range of potential landscapes and attractors.

\subsection{Example with a single fixed point}

\label{subsec_A}
Consider a two-dimensional example exhibiting a single fixed point at the origin. The drift
 ${b = -\nabla V + l}$ is given by
\begin{equation}
\label{eq:ex1_drift}
b(x,y) = \begin{bmatrix} -x^3 +y^3\\ - y^3 -x^3
\end{bmatrix},
\end{equation}
from which,
\begin{equation}
\label{eq:ex1_potential}
V(x,y) = \frac{x^4}{4} + \frac{y^4}{4} , \quad l = \begin{bmatrix}
y^3 \\
-x^3
\end{bmatrix},
\end{equation}
and satisfies the orthogonality condition $\inner{\nabla V}{l}=0$ (see equation \ref{eq:orthogonal}). 

\begin{figure}
\begin{center}
\includegraphics[width=246pt]{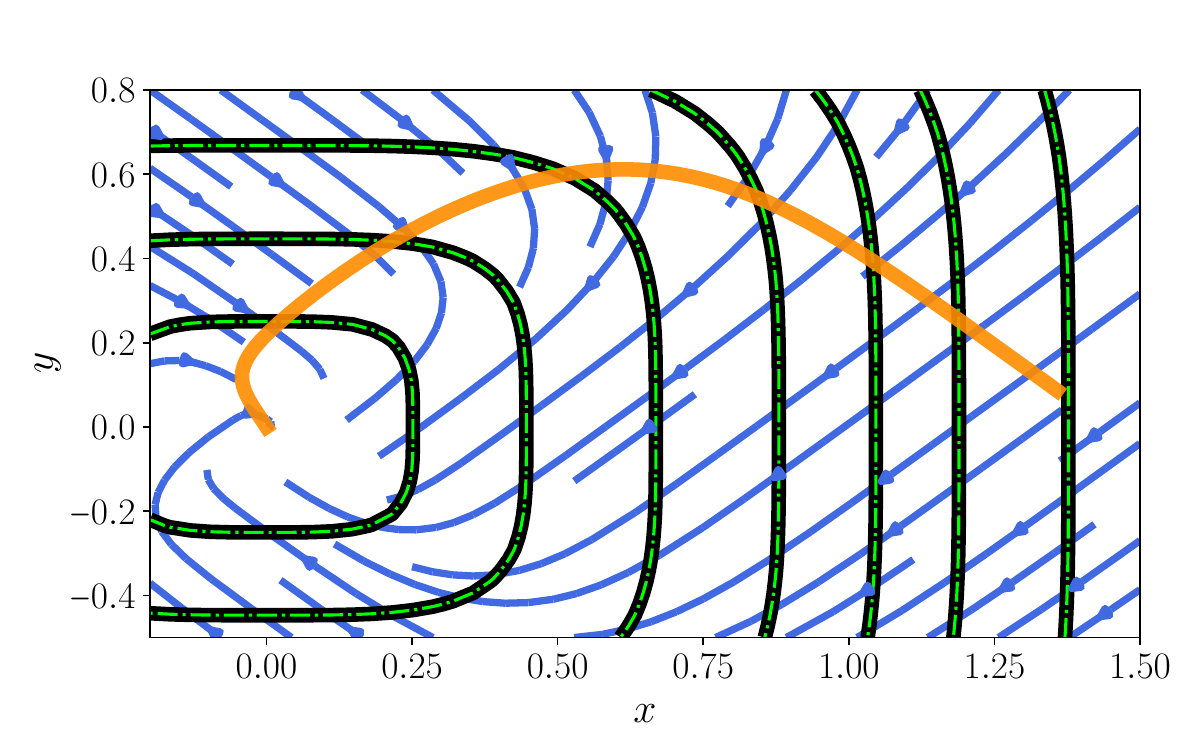}
\caption{Instanton trajectory beginning at $(0,0)$ and exiting the basin of the fixed point ending up at $(1.38,0.09)$ is shown in orange and plotted against the drift vector field (\ref{eq:ex1_drift}) in blue. The solid black lines represent the theoretical level lines of the quasi-potential (\ref{eq:ex1_potential}), while the dash-dot contour levels represent the computed quasi-potential obtained from SINDy after integration (\ref{eq:ex1_grad}). This comparison illustrates that both level curves are closely aligned.\label{fig:instanton_cubic}}
\end{center}
\end{figure}

\begin{figure*}
\begin{center}
\includegraphics[width=\linewidth]{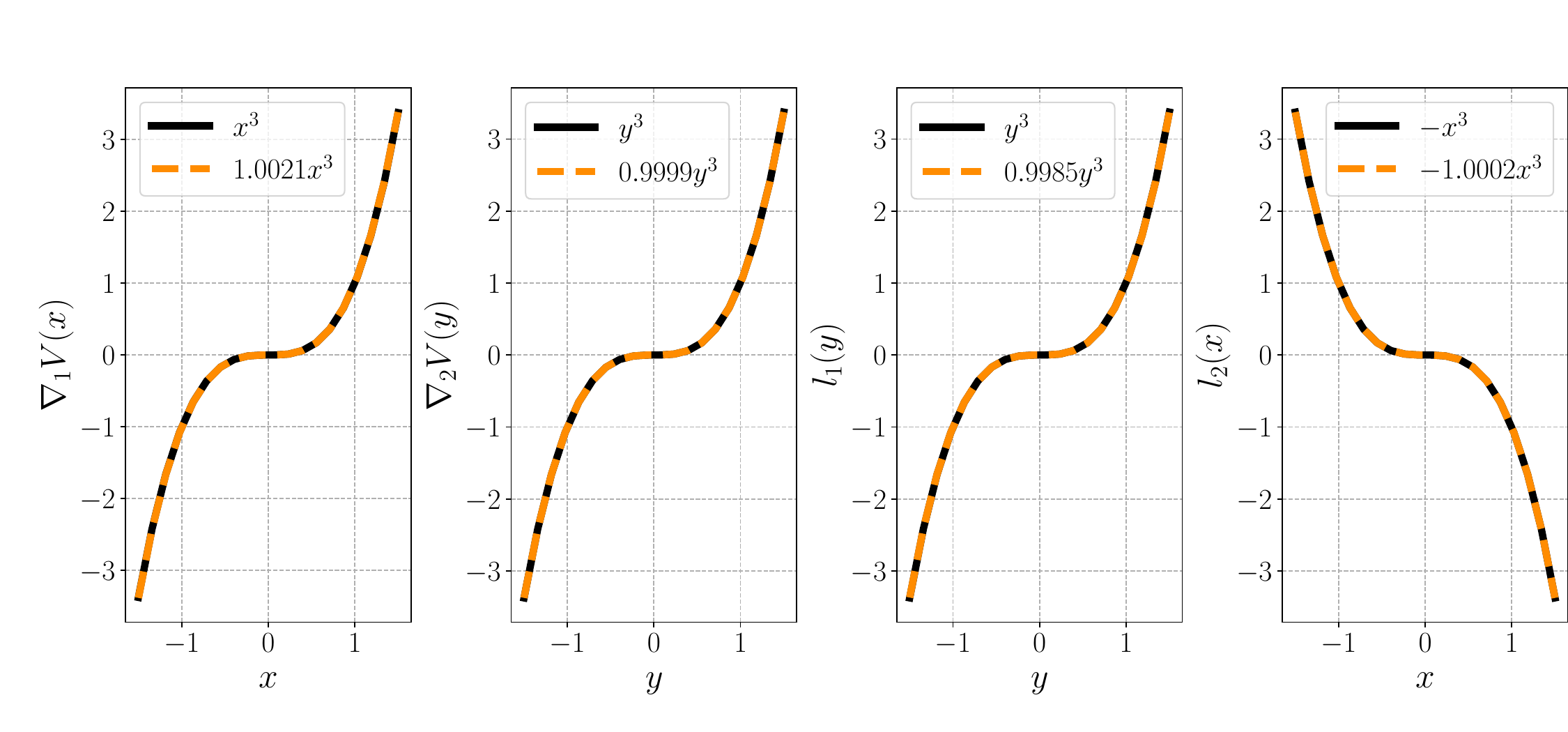}
\caption{The components of the drift (\ref{eq:ex1_drift}), gradient  $\nabla V$ and orthogonal $l$, are displayed. The solid lines correspond to theoretical values (\ref{eq:ex1_potential}), whereas the dashed lines depict the components learned by SINDy (\ref{eq:ex1_grad}-\ref{eq:ex1_l}). Both results exhibit perfect alignment.  
\label{fig:comps_cubic}}
\end{center}
\end{figure*}

We start by obtaining the instanton via solving Hamilton’s equations (\ref{eq:hamiltons_a}-\ref{eq:hamiltons_b}) with time reparametrization in accordance with  \cite{grafke-grauer-schaefer-etal:2014,grigorio:2020} and the mid-point rule for the Hamiltonian. The system is initiated from the attractor (the origin). Instanton $\varphi(t)$ starting at $(0,0)$ and ending up outside the basin of the fixed point is depicted (orange) together with the drift vector field $b(x,y)$ (blue) in Fig.~\ref{fig:instanton_cubic}. 

Next, SINDy allows us to estimate both drift components (gradient and normal) from the solution of Hamilton’s equations (\ref{eq:hamiltons_a}-\ref{eq:hamiltons_b}), the instanton $\varphi(t)$ along with $\vartheta(t)$. We constructed a library with functions ${\mX = [1\;\; x\;\; y\;\; x^2\;\; y^2\;\; xy\;\; x^3 \;\; y^3\;\; x^2 y\;\; xy^2 ]}$ as suggested by the form of the drift (\ref{eq:ex1_drift}), where $x$ and $y$ are column vectors. The results show that the computed gradient component is, 
\begin{equation}
\label{eq:ex1_grad}
\nabla V(x,y) = \begin{bmatrix}
0.9999\, x^3 \\
1.0021 \,y^3
\end{bmatrix},
\end{equation}
with a maximum relative error of $0.21\%$ compared to the theoretical field $\nabla V(x,y) = [x^3 \;\;y^3]^\transp$. Correspondingly, we found that the orthogonal component of the drift,
\begin{equation}
\label{eq:ex1_l}
l(x,y) = \begin{bmatrix}
0.9985 \,y^3 \\
-1.0002\, x^3
\end{bmatrix},
\end{equation}
is aligned with the theoretical vector $[y^3\; -x^3]^{\transp}$, resulting in a relative error of $0.15\%$. Further, once we have learned the gradient component (\ref{eq:ex1_grad}) from SINDy simulations, we can determine the quasi-potential through analytical integration.

These numerical findings are visually represented in Fig.~\ref{fig:instanton_cubic}. The computed quasi-potential derived from integrating SINDy result (\ref{eq:ex1_grad}), represented by dash-dot contour lines, is compared with the theoretical quasi-potential (\ref{eq:ex1_potential}), described as solid black lines, illustrating that the curves are well-matched. Fig.~\ref{fig:comps_cubic} shows the drift components, gradient $\nabla V$ and orthogonal $l$, as predicted by our method (\ref{eq:ex1_grad}-\ref{eq:ex1_l}) (dash-dot lines), with the theoretical components (\ref{eq:ex1_potential}) (black solid lines). The two findings show excellent agreement.

\subsection{3D System}

One of the advantages of sparse identification SINDy applied to the instanton framework to find the drift decomposition is that it can be easily extended to higher dimensions. This differs from previous methods used to evaluate the quasi-potential, where the numerical algorithm for 2D systems \cite{CAMERON:2012} cannot be directly applied to 3D systems, requiring the development of a new solver \cite{Cameron_etal:2019}. In contrast, our approach permits the evaluation of the quasi-potential within the same algorithm. This demonstrates that the 3D case can be seen as a natural extension of the 2D case. We emphasize that this scalability, i.e., the straightforward extension to higher dimensions, lies within the SINDy, a common feature of machine learning methods.

Let us examine the example provided in \cite{Cameron_etal:2019}, which is defined as:
\begin{equation}
\label{ex:3d_drift}
b(x,y,z) = \begin{bmatrix}
-3 & -4 & -1 \\
3 & -4 & -1 \\
3 & 4 & -1
\end{bmatrix} 
\begin{bmatrix}
x \\ y \\ z
\end{bmatrix},
\end{equation}
such that 
\begin{equation}
\label{ex:3d_l}
V(x,y,z) = \frac{3x^2}{2}+2y^2+\frac{z^2}{2},\quad l = \begin{bmatrix}
-4y-z\\ 3x-z\\ 3x+4y
\end{bmatrix}.
\end{equation} 

As in the previous example, we utilize Hamilton’s equations (\ref{eq:hamiltons_a}-\ref{eq:hamiltons_b}) in a time-reparametrized form \cite{grigorio:2020,grafke-grauer-schaefer-etal:2014} to determine the instanton $\varphi(t)$. We then apply sparse identification to obtain the drift orthogonal decomposition. Considering the form of $b(x,y,z)$, we carefully select candidate functions ${[1\;\; x\;\; y\;\;z\;\; x^2\; y^2\;\;z^2\;\; xy\;\;xz\;\;yz ]}$ to create the library matrix $\mX$. We obtained the gradient part,
\begin{equation}
\label{eq:ex2_grad}
\nabla V(x,y,z) = \begin{bmatrix}
3.0000\,x\\
4.0000\,y\\
1.0000\,z
\end{bmatrix}.
\end{equation}
In analyzing the gradient component, we observed that (\ref{eq:ex2_grad}) exhibited a relative error lower than $0.22\times 10^{-6}$ when compared with the theoretical gradient of (\ref{ex:3d_l}). Meanwhile, our findings for the transverse component $l(x)$ yielded,
\begin{equation}
l(x,y,z) = \begin{bmatrix}
-4.0000\,y-1.0000\,z \\
3.0000\,x-1.0000\,z\\
3.0000\,x+4.0000\,y
\end{bmatrix}.
\end{equation}
It shows a relative error of $0.025\%$ when compared to its theoretical value (\ref{ex:3d_l}). Both significantly low relative errors indicate the accuracy of the adopted approach. 

\subsection{Bistable system}
\label{subsec_bistable}

So far we have considered examples with a single stable fixpoint. Now, we present an example where the vector field $b$ exhibits bistability, i.e., two stable fixpoints. The model is drawn from \cite{BR:2022}, defined by,
 \begin{equation}
 \label{ex:bystable_V}
 V(x,y) = \frac{x^4}{4} - \frac{x^2}{2} + \alpha \, \frac{y^2}{2}\,, 
 \end{equation}
 and transverse field,
 \begin{equation}
 \label{eq:bystable_l}
 l(x,y) = \beta \, x \begin{bmatrix}
 -\alpha \, y \\
 x^3 - x
 \end{bmatrix},
 \end{equation}
where $\alpha$ and $\beta$ are constants. Without loss of generality, our simulation utilizes $(\alpha,\beta) = (2,1)$. It is straightforward to check that the orthogonality condition $\inner{\nabla V}{l}=0$ is met. The critical points of this system are $(-1,0)$, $(0,0)$ and $(1,0)$, of which $(-1,0)$ and $(1,0)$ are stable fixpoints, while $(0,0)$ has one unstable direction.

\begin{figure}
\begin{center}
\includegraphics[width=246pt]{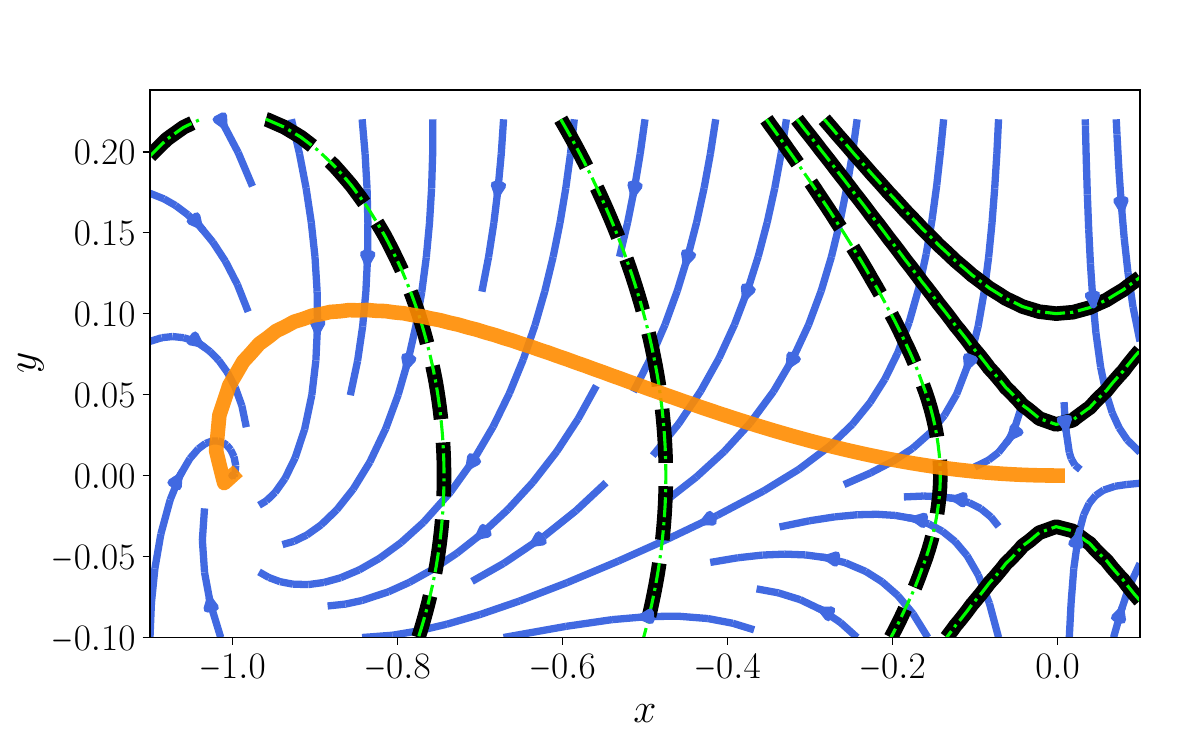}
\caption{Instanton (orange) transitioning from attractor at $(-1,0)$ to the saddle $(0,0)$ plotted against drift vector field (blue). Black lines represent the theoretical level lines of the quasi-potential. Green dash-dot contour lines represent the predicted quasi-potential obtained from SINDy after integration of (\ref{eq:ex_bistable_grad}). Both quasi-potential curves exhibit consistent agreement. \label{fig:instanton_bistable}}
\end{center}
\end{figure}

Unlike the previous examples, the instanton cannot be treated with a CS-like algorithm as the quasi-potential is non-convex \cite{alqahtani-grafke:2021}. To find the maximum likelihood transition path (instanton), we apply the (simplified) geometric minimum action method (GMAM) \cite{GSE:2017,grafke-vanden-eijnden:2019}. Fig.~\ref{fig:instanton_bistable} depicts the instanton transition from the attractor at $(- 1,0)$ to the saddle at $(0,0)$ in orange, plotted against the drift vector field ${b = -\nabla V + l}$ (blue).

\begin{figure*}[htp]
\includegraphics[width=\linewidth]{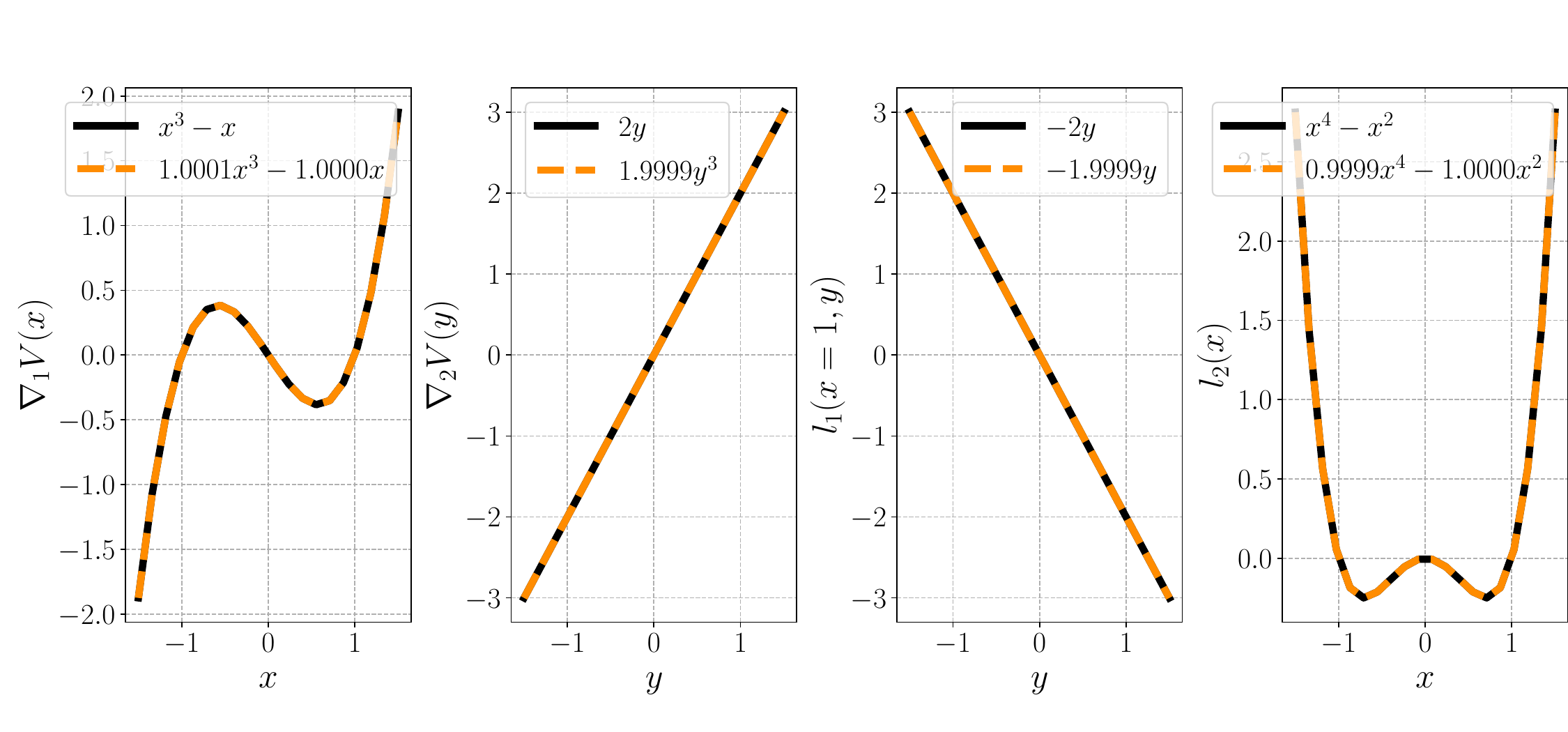} %
\caption{The drift components $\nabla V$ and $l$ for \ref{subsec_bistable}. Solid lines show theoretical values (the gradient of (\ref{ex:bystable_V}) and (\ref{eq:bystable_l})), accurately matched with dashed lines representing predicted components by SINDy, equations (\ref{eq:ex_bistable_grad}-\ref{eq:ex_bistable_l}).\label{fig:comps_bistable}}
\end{figure*}

The SINDy algorithm was applied to the instanton results as input, using a library of terms ${[1\;\; x\;\; y\;\; x^2\;\; y^2\;\; xy\;\; x^3 \;\; y^3\;\; x^2 y\;\; xy^2 \;\; x^4\;\; y^4\\ \;\; x^3y\;\;x^2y^2\;\;xy^3]}$. It produced equations,
\begin{equation} % align}
\label{eq:ex_bistable_grad}
\nabla V(x,y) = \begin{bmatrix}
 1.0001\,x^3-1.0000\,x\\
1.9999\,y
\end{bmatrix}, \\ 
\end{equation} % align}
and,
\begin{equation}
\label{eq:ex_bistable_l}
l(x,y) = x\begin{bmatrix}
 -1.9999\,y \\
 (0.9999\,x^3 - 1.0000\,x)
\end{bmatrix}.
\end{equation}
Importantly, the most significant relative error observed was less than $1\%$. These components estimated by SINDy are visually represented in Fig.~\ref{fig:comps_bistable}, allowing for a direct comparison with the theoretical components, the gradient of (\ref{ex:bystable_V}) and (\ref{eq:bystable_l}), and demonstrating excellent consistency. In addition, Fig.~\ref{fig:instanton_bistable} portrays both the learned and theoretical quasi-potentials. The black lines represent the theoretical level lines of the quasi-potential, while the light dash-dot green contour lines represent the predicted quasi-potential produced by SINDy after integrating (\ref{eq:ex_bistable_grad}). It shows that these curves are appropriately aligned.

\subsection{Limit cycle}
\label{subsec:limit_cycle}

Consider the following model: The attractor is a limit cycle,
\begin{equation}
\label{eq:ex_cycle_drift}
b(x,y) = \begin{bmatrix}
-x(x^2+y^2-1) +y\\
-y(x^2+y^2-1)-x
\end{bmatrix},
\end{equation}
with,
\begin{equation}
\label{eq:ex_limit_cycle_pot}
V(x,y) = \frac{1}{4}(x^2+y^2)^2 - \frac{1}{2}(x^2+y^2)\,,\quad l(x,y)=\begin{bmatrix}
y \\
-x
\end{bmatrix}.
\end{equation}

\begin{figure}[htbp]
\begin{center}
\includegraphics[width=246pt]{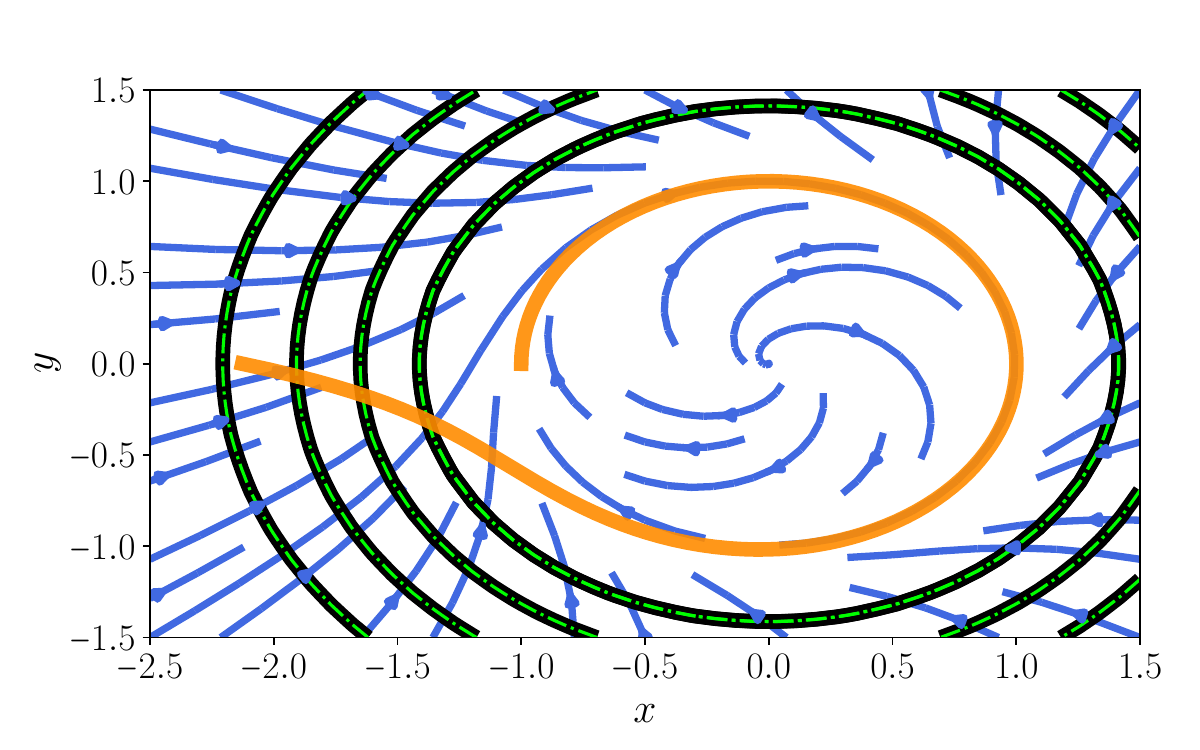}
\end{center} 
\caption{Instanton (orange) plotted against drift vector field (blue). The system is initialized within the limit cycle at point $(-1,0)$. It travels along the limit cycle following the orthogonal component $l$ until it reaches a point where it ascends the potential, eventually leaving the attractor and ending up at point $(-2.18,0.0)$. Theoretical (black lines) and predicted (green dash-dot lines) quasi-potential contour levels.\label{fig:instanton_cycle}}
 \end{figure}

Instanton simulation has been initiated within the limit cycle. Fig.~\ref{fig:instanton_cycle} presents the instanton trajectory beginning at limit cycle connecting the points $(-1,0)$ and $(-2.18,0.0)$. It is noticed that the system runs along the limit cycle until an instant when it departs the attractor.

In the next step, SINDy procedure with the library of terms ${[1\;\; x\;\; y\;\; x^2\;\; y^2\;\; xy\;\; x^3 \;\; y^3\;\; x^2 y\;\; xy^2 \;\; x^4\;\; y^4\\ \;\; x^3y\;\;x^2y^2\;\;xy^3]}$, produced, 
\begin{equation}%align}
\label{eq:ex_limit_cycle_grad}
\nabla V(x,y) = \begin{bmatrix}
 1.0000\,x\,(-1+x^2+y^2)\\
1.0000\,y\,(-1+x^2+y^2)
\end{bmatrix}, \\
\end{equation}%align}
and,
\begin{equation}%align}
\label{eq:ex_limit_cycle_l}
l(x,y) = \begin{bmatrix}
 1.0000 \, y \\
 - 1.0000 \, x
\end{bmatrix}.
\end{equation}
When considering accuracy until the fourth decimal place, there is a full alignment between analytical and predicted quasi-potential.
It should be noted that, contrary to the previous cases, we cannot obtain the quasi-potential if a fifth decimal place is considered. In this case, each term in equation (\ref{eq:ex_limit_cycle_grad}) gets a different factor. As a result, the mixing of $x$ and $y$ in the equation of $\nabla V$ (\ref{eq:ex_limit_cycle_grad}), prevents the identification of a scalar whose gradient matches the form in (\ref{eq:ex_limit_cycle_grad}). Thus, writing equation (\ref{eq:ex_limit_cycle_grad}) with $\nabla V$ is an abuse of notation; however, by approximating the coefficients to the fourth decimal place, the difference between the factors in equation (\ref{eq:ex_limit_cycle_grad}) disappears, allowing us to obtain the quasi-potential aligned with the theoretical value in equation (\ref{eq:ex_limit_cycle_pot}), as depicted in Fig. \ref{fig:instanton_cycle} where  theoretical (black lines) and predicted (green dash-dot lines) quasi-potential have matching contour levels.

\section{Conclusion}
\label{Sec:Conclusion}

Our study demonstrates a successful application of the sparse identification of nonlinear dynamics (SINDy) along with instanton method to construct a global landscape of the quasi-potential. The minimum action path (instanton) encodes information about the drift components: The quasi-potential gradient and orthogonal parts. SINDy is capable of finding these components from instanton, providing an analytical expression for the drift decomposition $b(x) = -\nabla V(x)+l(x)$, from which quasi-potential can be easily obtained. We emphasize that this approach is pertinent to both two-dimensional and easily extended three-dimensional applications and is utilized across a range of scenarios. We leave for future investigations the case of stochastic partial differential equations and multiplicative noise. Concerning the latter, a preliminary study revealed bigger errors in estimating drift components compared with the additive noise case, suggesting a more detailed study about the state dependent noise case.

Typically, obtaining a numerical approximation of the quasi-potential within the instanton method entails the computation of numerous trajectories, incurring significant costs. However, our approach employs sparse identification based on a single trajectory, instanton. Moreover, our method presents an advantage over techniques utilizing machine learning with neural networks to determine the quasi-potential \cite{Li_etal:2021,Lin_etal:2021, Li_etal:2022} in that usually many trajectories have to be used in the training phase. Within the SINDy method, only the instanton path is required for learning the drift decomposition and quasi-potential.

Additionally, our work highlights that the instanton is not only paramount in characterizing extreme events and transition pathways but also is important as it carries information about the global quasi-potential which can be retrieved with the help of SINDy. Besides, our findings yields the orthogonal component of the drift, relevant to the evaluation of next to leading order contributions to the decay rate of statistical quantities as transition probabilities, first exit times, and mean first passage times.

\bibliography{bib}

\begin{appendices}

%\appendix
\section{Hamilton-Jacobi (HJ)}\label{sec:app}
Here, we review the decomposition of the vector field $b$ into potential and rotational components. Start with the time-dependent HJ equation. Let $x \in \R^d$, $T \in \R$, and $\pol$ be a scalar such that the  probability distribution follows the Wentzel-Kramers-Brillouin (WKB) ansatz \cite{applied:2019},
\begin{equation}
\label{A1}
p(x,T|\bar{x}) = A \expp{-2 \, \pol}{\vep}.
\end{equation}
By substituting this ansatz into the Fokker-Planck equation,
\begin{equation}
\label{A2}
\p_t \, p(x,T|\bar{x}) = -\nabla_i (b_i \, p(x,T|\bar{x})) + \frac{\vep}{2} \,\nabla_i \nabla_j (\sigma_i \, \sigma_j \, p(x,T|\bar{x})),      \quad \,   i,j = 1,\dots, d,
\end{equation}
(Einstein summation implied) and  grouping leading-order terms, we find,
\begin{equation}
\label{A3}
-\p_t \,  \pol =  \, \inner{b}{\nabla  \pol } +  \inner{\nabla  \pol }{a \, \nabla  \pol },
\end{equation}
where $a(x) = \sigma \sigma^{\transp}(x)$ is the diffusion matrix. This equation is known as the HJ equation \cite{applied:2019}. Assuming stationarity, $\p_t \,  \pol  = 0$, equation (\ref{A3}) takes the form,
\begin{equation}
\label{A4}
\inner{b}{\nabla \pol} +  \inner{\nabla \pol}{a \, \nabla \pol} = 0.
\end{equation}
For $\nabla \pol \neq 0$, equation (\ref{A4}) invites the decomposition of $b(x)$ as follows,
\begin{equation}
\label{A4a}
b(x) = -  a\, \nabla \pol+ l(x), \,
\end{equation}
where $l(x)$ satisfies, 
\begin{equation}
\label{A5}
\inner{ l }{\nabla \pol} = 0,  
\end{equation}
which shows that the field $l(x)$ is orthogonal to the gradient of the quasi-potential. On the other hand, in the non-stationary case, equation (\ref{A4a}) still holds provided that, 
\begin{equation}
\label{A6}
 \inner{\nabla \pol}{l} = - \p_t \, \pol,
\end{equation}
indicating that the orthogonality between $l$ and $\nabla \pol$ is no longer preserved if stationarity is not fulfilled.

\section{The stepwise sparse regressor}
\label{sec:app_2}

Here are more details about the stepwise sparse regressor SSR \citep{BFC:2018}, an adaptation of sparse identification of non-linear dynamics (SINDy) \cite{BPK:2016} used in the paper. As discussed in \ref{sec:SINDy}, given the datasets $y_j(t_1),\dots,y_j(t_N)$ and $x_j(t_1),\dots,x_j(t_N)$, $j=1,\dots,d$, we admit $y=F(x)$ and search for a model $F$ that describes the datasets accurately with minimum complexity. In other words, after organizing the datasets according to (\ref{eq:Y}-\ref{eq:YXc}\ref{eq:library}), that is, writing $y$ as a linear combination basis functions of $x$,  $\mY = \mX c$, we are seeking a sparse representation of the matrix $c$. However, when imposing a sparse solution, we have to bear in mind that an excessively sparse $c$ will not give an accurate representation (under-fitting). In contrast, a barely sparse $c$ will tend to overfit the data. There is a trade-off between under and over-fitting that we are aiming at. In SSR \cite{BFC:2018}, sparsity is enforced iteratively by removing one element of $c$ at each step. The SSR scheme is outlined in algorithm \ref{alg}. Then, cross-validation is used to select the solution of optimal sparsity.

In a nutshell, SSR comes down to the following steps:
\begin{itemize}
\item Compute a standard least squares regression
\begin{equation}
\label{eq:A2_lsq}
\tilde{\mathbf{c}}_j = \underset{\mathbf{c}_j\in \R^K}{\mathrm{argmin}}\,\bigl( y_j-\mX \mathbf{c}_j \bigr)^{\mathrm{T}}\bigl( y_j-\mX \mathbf{c}_j \bigr)\,,\;\;j=1,\dots,d.
\end{equation}

\item Set the smallest coefficient of $\tilde{\mathbf{c}}_j$ to zero, that is, 
%\begin{equation}
%\label{eq:A2_}
%\mathbf{c}_j(i) = 0, \;\; i =\underset{k}{\mathrm{min}}\,\tilde{c}_{kj} \,.
%\end{equation}
\begin{equation}
\label{eq:A2_}
\mathbf{c}_j(i) = 0, \;\; i =\underset{k}{\mathrm{argmin}}\,\tilde{c}_{kj} \,.
\end{equation}

\item Perform the least squares regression on the remaining degrees of freedom. That is, apply regression to
\begin{equation}
\label{eq:A2_lsq_hat}
\mY_{nj} = \sum_{k\neq i}^K \mX_{nk} \, c_{kj} \,,
\end{equation}
with $i^{\text{th}}$ column of $\mX$ removed, and correspondingly, the $i^{\text{th}}$ row of $c$ excluded.
\item Repeat until all coefficients vanish.
\item Apply cross-validation to select the optimal level of sparsity.
\end{itemize}

\begin{figure}
\includegraphics[width=\linewidth]{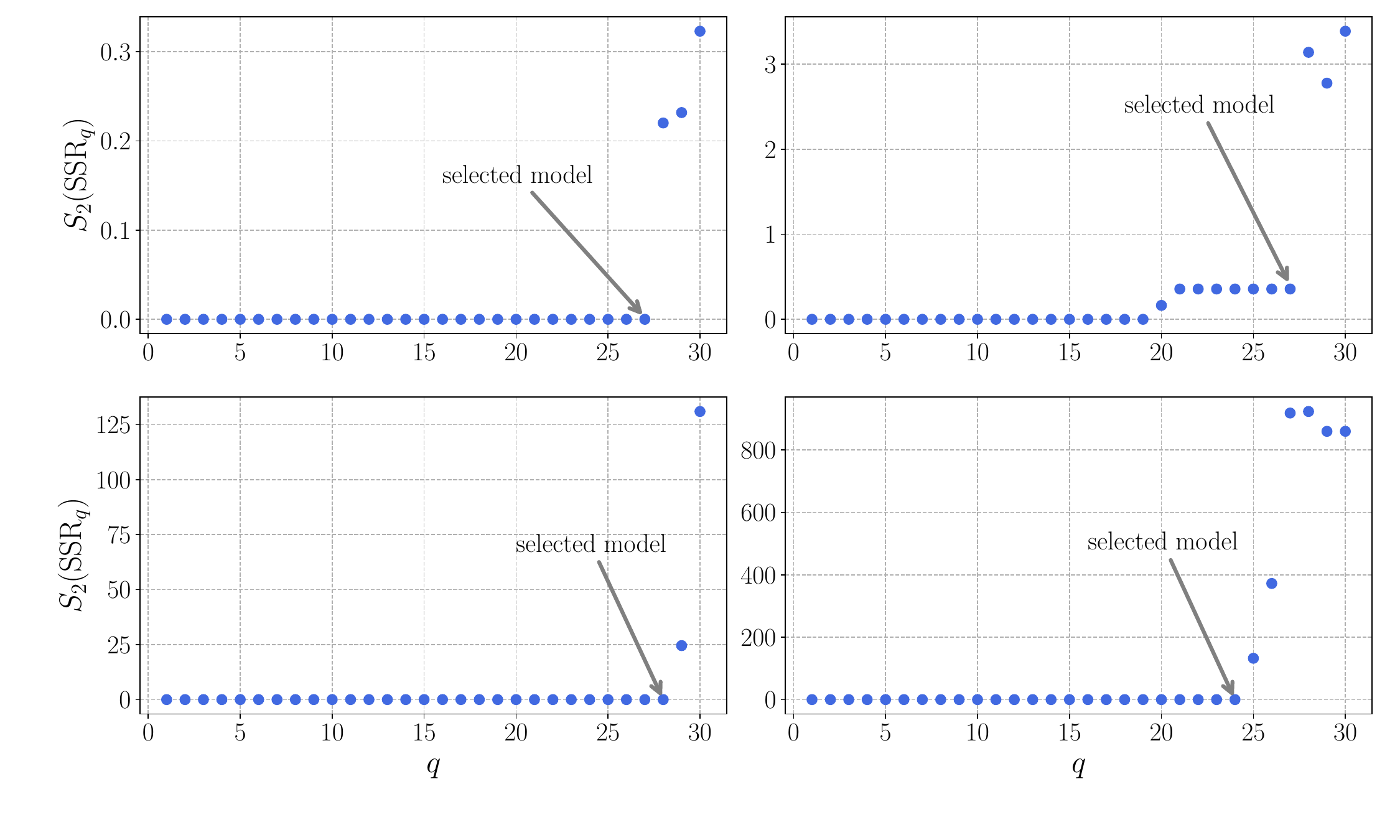}
\caption{\label{fig:A2_cv}Cross validation score $S_2(\mathrm{SSR})_q$ as function of sparsity $q$. Left: The orthogonal component. Right: The gradient component. Upper panel: Bistable problem \ref{subsec_bistable}. Bottom panel: Limit cycle problem \ref{subsec:limit_cycle}. The notable surge in the score for a large $q$ indicates that a too-sparse representation is unable to capture the key aspects of the system. Therefore, the selected model has the highest level of sparsity, where the arrow is pointing, just before the shape starts to increase.}
\end{figure}

{
\setlength\arrayrulewidth{1.5pt}
\begin{center}
\begin{tabular}{l}
\label{alg}
{\bf \textsf{Algorithm 1.}} SSR algorithm \\ 
\hline
{\bf Choose:} \\
$k < N$ \quad \# number of folds\\
{\bf Set:}\\
$q=0$ \quad \# sparsity level\\
$p=1$\\
$r = N/k$\\
{\bf While} $p\leqslant k$\\
\hspace{1cm} $t_V = \{t_{(p-1)r+1},\dots,t_{pr}\},\; t_T=\{t_{1},\dots,t_N\}-t_V$ \\
\hspace{1cm} $c \leftarrow$ least squares $(\mY_{t_T}=\mX_{t_T}c)$\\
\hspace{1cm} $i,j = \underset{m,n}{\mathrm{argmin}}\;c[m,n]$ \\

\hspace{1cm} {\bf While} $q \leqslant (K\cdot d)$\\
\hspace{2cm} $c[i,j] = 0$\\
\hspace{2cm} $zeroinds = \mathrm{zeros}(c)$ \quad \# indices for which $c$ is zero\\
\hspace{2cm} {\bf While} $col \leqslant d$ \\
\hspace{3cm} $\hat{i} = NOT(zeroinds[:,col])$\\
\hspace{3cm} $c[\hat{i},col] \leftarrow$ least squares $(\mY_{t_T}[:,n]=\mX_{t_T}[:,\hat{i}]\,c[\hat{i},col])$ \\
\hspace{3cm} $col \leftarrow col+1$ \\
\hspace{2cm} $i,j = \underset{m,n}{\mathrm{argmin}}\;c[m,n]$ \\
\hspace{2cm} $s_2[l,q]= \mathrm{Tr}\left[ \bigl(\mY_{t_V}-\mX_{t_V}  c \bigr)^{\mathrm{T}}\bigl(\mY_{t_V} - \mX_{t_V}  c \bigr) \right]$ \\
\hspace{2cm} $q \leftarrow q+1$ \\
\hspace{1cm}$p \leftarrow p+1$ \\
$S_2[q] = \frac{1}{k}\sum_{p=1}^{k}s_2[p,q]$ \# average error \\
{\bf Terminate} \\
 \hline

\end{tabular} 
\end{center}
}

Cross-validation is a widely used approach for model selection \cite{statisticallearning}. One of its variants is the so-called $k$-fold cross-validation, which uses part of the data to fit the model and a different part to test it. Particularly, the data set is split into $k$ disjoint subsets (folds). The first fold is treated as a control set (left for validation), and the model is fit into the remaining $k-1$ folds; that is, the $k-1$ folds are used as training sets. Then, this procedure is repeated $k$ steps, such that at each step a different fold is treated as control data. As a result, we end up with $k$ estimates of the predicted error, one for each fold tested against the control. Averaging over these errors gives the score, a function used to evaluate the predictive error of each model.

At each iterative step a set of coefficients $c$ is generated with 
a different level of sparsity. For instance, after $q^{\text{th}}$ iteration, a matrix $c$ with $q$ elements equal to zero is produced. We shall denote the solution $c$ obtained from the algorithm on $q$ iterations as $\mathrm{SSR}_q$, $q=1,\dots,K \times d$. To select among the several models $\mathrm{SSR}_q$, $k$-fold cross-validation criterion is considered ($k=10$ was utilized in our case). Let $A_i$, $i=1,\dots,k$ represent the fold used for validation and $B_i = \cup_{j \neq i}  A_j$ ($j=1,\dots,k$) is the $k-1$ folds used for training. Then, a score function quantifying the predicted error  $S_2(\mathrm{SSR}_q)$ for each model $\mathrm{SSR}_q$ is defined by
\begin{align}
\label{eq:A2_cv}
S_2(\mathrm{SSR}_q) = \frac{1}{k}\sum_{i=1}^{k} \mathrm{Tr}&\left[\bigl(\mY_{A_i}-\mX_{A_i} \cdot \mathrm{SSR}(\mX_{B_i},\mY_{B_i})_q\bigr)^{\mathrm{T}} \,\bigl(\mY_{A_i}-\mX_{A_i} \cdot \mathrm{SSR}(\mX_{B_i},\mY_{B_i})_q\bigr)\right],\nonumber \\ 
& \hspace{2cm}B_i = \underset{\hspace{-16pt}j\neq i}{\bigcup^{k}_{j=1} \; A_j}\,,
\end{align}
where $\mathrm{SSR}(\mX_{B_i},\mY_{B_i})_q$ is the $q$-sparse coefficients $c$ obtained with the use of the subset $B_i$ for training.

Typically, when the score $S_2(\mathrm{SSR})_q$ is plotted against the sparsity level $q$, there is a sudden increase for large enough $q$, as shown in Fig.~\ref{fig:A2_cv}. This means that an overly sparse representation is insufficient to capture the essential features of the system. Finally, the criterion for selecting the model was to pick the maximum level of sparsity before the score displays a sharp climb as indicated in Fig.~\ref{fig:A2_cv}. This rationale was also taken into account in \cite{callaham:2021}, even though a different function was used for the score. 
\end{appendices}

\end{document}